\documentclass[]{article}
\usepackage{amsmath,amssymb}
\usepackage[a4paper]{geometry}
\usepackage{graphicx}
\usepackage{color,float}
\usepackage{subfig}

\title{ Thermodynamics of Rotating Kaluza-Klein Black Holes in Gravity's Rainbow}
\author{Salwa Alsaleh}
 \date{ Department of Physics and Astronomy,\\ King Saud University, Riyadh 11451, Saudi Arabia
 \\ \today}

\begin{document}

\maketitle

\begin{abstract}
In this paper, a four dimensional  rotating Kaluza Klien (K-K) black hole was deformed using rainbow functions derived from loop quantum gravity and non-commutative geometry. We studied the thermodynamic properties and critical phenomena of this deformed black hole. The deformed temperature and entropy showed the existence of a Planckian remnant.  The calculation of Gibbs free energy $ G$ for the ordinary and deformed black holes showed that both share a similar critical behaviour. 
\end{abstract}
\section{Introduction}
The quest for a consistent theory of gravity is on going since the early 20's in the past century. Nevertheless, such a theory is still unavailable. Many programmes for quantum gravity however exists, like string theory, loop quantum gravity (LQG), causal dynamical triangulation (CDT), and many others. Most of these programmes prridect that the spacetime admits a minimal length scale. Therefore , there is a maximal energy $ E_P$ that can be put into a system. This basic, yet important  and universal prediction of quantum gravity programmes leads to phenomenological investigation of quantum gravity.  The Ho\v{r}ava-Lifshitz gravity is based on such investigation, by imposing a deformation to the energy-momentum dispersion relations for energies close to $ E_p$ \cite{hovrava2009quantum,hovrava2009spectral}. Another deformation is made by gravity's rainbow \cite{magueijo2002lorentz}, where different wavelengths of light ( having different energies) experience gravity differently. More generally, gravity is energy-dependent phenomena. \\ The deformation of energy-momentum dispersion relations can be derived from different quantum gravity programmes, in the UV limit . For instance in spacetime
foam \cite{amelino1998tests}, spin-network in loop quantum gravity (LQG) \cite{Gambini:1998it}, discrete spacetime
\cite{tHooft:1996ziz}, models based on string field theory \cite{Kostelecky:1988zi} and non-commutative geometry \cite{carroll2001noncommutative}.
This formalism has been heavily studied within string theory as well, the different Lifshitz scaling of space and time has been used to deform 
type IIA string theory \cite{gregory2010lifshitz}, type  IIB string theory \cite{Burda:2014jca},   AdS/CFT correspondence 
\cite{Gubser:2009cg, Ong:2011km, Alishahiha:2012iy,  dey2015interpolating},  dilaton black branes 
\cite{Goldstein:2010aw, bertoldi2009thermodynamics}, and dilaton black holes \cite{Zangeneh:2015uwa, Tarrio:2011de}.\\
It has been shown that  Ho\v{r}ava-Lifshitz gravity  and gravity's rainbow produce similar physical results\cite{Garattini:2014rwa}, as they are based on the same physical assumption.The Lifshitz deformation of geometries has produced interesting results, 
and rainbow deformation has the same motivation, 
in this paper we will study the rainbow deformation of rotating Kaluza-Klien black holes. 
In gravity's rainbow, the geometry depends on 
the energy of the  probe, and thus  probes of 
of different energy see the geometry  differently.
Thus,  a single metric is replaced  by a family of energy
dependent metrics forming a rainbow of metrics. Now the 
UV modification of the energy-momentum dispersion relation  can be expressed as 
\begin{equation}  \label{MDR}
E^2f^2(E/E_P)-p^2g^2(E/E_P)=m^2
\end{equation}
where $E_P$ is the Planck energy, $E$ is the energy at which the geometry 
is probed, and  $f(E/E_P)$ and $g(E/E_P)$
are the  rainbow functions. As the general relativity should be recovered 
in the IR limit, we have 
\begin{equation}
\lim\limits_{E/E_P\to0} f(E/E_P)=1,\qquad \lim\limits_{E/E_P\to0} g(E/E_P)=1.
\label{rainbowfunctions}
\end{equation}
Now the metric in gravity's rainbow \cite{magueijo2004gravity}
\begin{equation}  \label{rainmetric}
h(E)=\eta^{ab}e_a(E)\otimes e_b(E).
\end{equation}
So, the energy dependent   frame fields  are 
\begin{equation}
e_0(E)=\frac{1}{f(E/E_P)}\tilde{e}_0, \qquad e_i(E)=\frac{1}{g(E/E_P)}\tilde{%
	e}_i. 
\end{equation}
Here $\tilde{e}_0 $ and $ \tilde{e}_i$ are the original energy independent frame fields. 
The deformation of geometry by the rainbow functions has been studied extensively, such as the study of black rings \cite{Ali:2014yea}, black branes  \cite{ashour2016branes}, higher dimensional microscopic black holes and the consequences of gravity's rainbow on their detection at the TeV scale at the LHC  \cite{Ali:2014qra}. Gravity's rainbow has also been used to address the black hole information paradox \cite{Ali:2014cpa Gim:2015zra,ali2015gravitational}, and in alternative theories of gravity \cite{Hendi:2016dmh,hendi2016charged, Hendi:2015hja, hendi2016charged, hendi2016charged, Rudra:2016alu,Hendi:2017pld,  Garattini:2012ec}. \\
In this paper, we shall study deformed rotating Kaluza-Kleinblack hole by the rainbow functions, and investigate its thermodynamic properties.  We start by a review of rotating Kaluza-Klein black holes, and their thermodynamics, then we deform the metric via the rainbow functions and discuss the implications of this deformation on the thermodynamics of these black holes.
\section{ Thermodynamics of rotating Kaluza-Klein black holes }
Kaluza-Klein black holes are a 5d uplifted solution of rotating black holes with electric $Q$ and magnetic $P$ charges \cite{gibbons1986black,rasheed1995rotating,larsen2000rotating}. This is a general solution to the dyonic solution (where $Q=P$). This solution is considered from the 4d Einstein-Maxwell-dilaton theory \cite{matos1997stationary}, or as a rotating D0-D6 bound state in string theory \cite{itzhaki1998d6+}. The rotating KK black hole contain a 4D dyonic Reissner-Nordstr\o{}m black hole and Myers-Perry black hole \cite{azeyanagi2009holographic}.  The KK solution in 5d pure Einstein gravity has the following metric:
\begin{equation}
ds_{(5)}^2 = \frac{H_2}{H_1}\left( R d\hat y+ A\right) ^2-\frac{H_3}{H_2}\left( d\hat t+B\right) ^2+H_1\left( \frac{d\hat r^2}{\Xi}+d\theta^2+ \frac{\Xi}{H_3}\sin^2\theta d\phi\right) 
\end{equation}
Where:
\begin{subequations}
\begin{align}
H_1 &= \hat r^2 +\mu^2 j^2 \cos^2 \theta + \hat r (p=2 \mu)+\frac{1}{2}\,\frac{p}{p+q}\,\left( p-2\mu\right) \left( q-2\mu\right) \nonumber \\
\qquad&+ \frac{1}{2}\frac{p}{p+q} \sqrt{(p^2-4\mu^2)(q^2-4\mu^2)}j \cos \theta.\\
H_2 &= \hat r^2 + \mu^2 j^2 \cos^2 \theta +\hat r(q-2 \mu)+\frac{1}{2}\, \frac{q}{p+q}(p-2\mu)(q-2\mu) \nonumber \\
\qquad& -\frac{1}{2}\frac{p}{p+q} \sqrt{(p^2-4\mu^2)(q^2-4\mu^2)}j \cos \theta. \\
H_3 &= \hat r^2 +\mu^2j^2 \cos^2 \theta - 2 \mu \hat r.\\
\Xi &= \hat r^2 + \mu^2j^2-2 \mu \hat r.
\end{align}
\end{subequations}
And:
\begin{align}
A &= \left[ \sqrt{\frac{q(q^2-4\mu^2)}{p+q}}\left( \hat r+ \frac{p-2 \mu}{2}\right) -\frac{1}{2}\sqrt{\frac{q^3(p^2-4 \mu^2)}{p+q}}\, j \cos \theta \right]  H_2^{-1}d\hat t \nonumber \\
\qquad&+ \Bigg[ - \sqrt{\frac{q(q^2-4\mu^2)}{p+q}} \, ( H_2+ \mu^2j^2 \sin^2 \theta)\cos \theta   \nonumber  \\
\qquad&  +\frac{1}{2}\sqrt{\frac{q(q^2-4\mu^2)}{p+q}} \left\lbrace p\hat r-\mu(p-2 \mu )+ \frac{q(q^2-4\mu^2)}{p+q}\right\rbrace \, j \sin^2 \theta\Bigg] H_2 ^{-1} d \phi\\
B &= \frac{1}{2} \, \sqrt{pq}\, \frac{{pq+4\mu^2}\hat r-\mu (p-2 \mu)(q-2 \mu)}{p+q}\, H_3 ^{-1} j \sin^2\theta d\phi
	\end{align}
With $R$ being the radius of the compactified fifth K-K dimension $ \hat y$ with the condition $ \hat y = \hat y + 2 \pi$. We can obtain the 4-D metric after the K-K reduction of $ \hat y$. 
\begin{equation}
ds^2 {(4)} = - \frac{H_3}{\sqrt{H_1 H_2}} \left( d\hat t+B \right)^2+\sqrt{H_1H_2 } \left( \frac{d \hat r^2}{\Xi}+ d\theta^2+ \frac{\Xi}{H_3}\sin^2 \theta d\phi\right) .
\end{equation}
There are four physical parameters that characterises the rotating K-K black hole, the mass $M$, electric and magnetic charges $ Q, P$ and the angular momentum $J$. They are given in terms of the parameters $ \mu, q,p$ and $ j$ :
\begin{subequations}
	\begin{align}
M &= \frac{p+q}{4} \label{mass} \\
Q &= \frac{1}{2}\left( \frac{q(q^2-4 \mu^2)}{p+q}\right) ^{1/2} \label{Q} \\
P &=  \frac{1}{2}\left( \frac{p(p^2-4 \mu^2)}{p+q}\right) ^{1/2} \label{P} \\
J&= \frac{\sqrt{pq}(pq+4 \mu^2)}{4 (p+q)} \, j \label{J}
	\end{align}
\end{subequations}
The K-K black hole has an event horizon at $ \Xi=0$,
\begin{equation}
r_{\pm} = \mu ( 1 \pm \sqrt{1- j^2})
\label{horizon}
\end{equation}
The Hawking temperature is then:
\begin{equation}
T_{0} = \frac{\mu  \hbar }{\pi  \sqrt{p q} \left(\frac{2 \mu }{\sqrt{1-j^2}}+\frac{4 \mu ^2+p q}{p+q}\right)}
\label{tzero}
\end{equation}
\begin{figure}[h]
	\centering
	\includegraphics[scale=1]{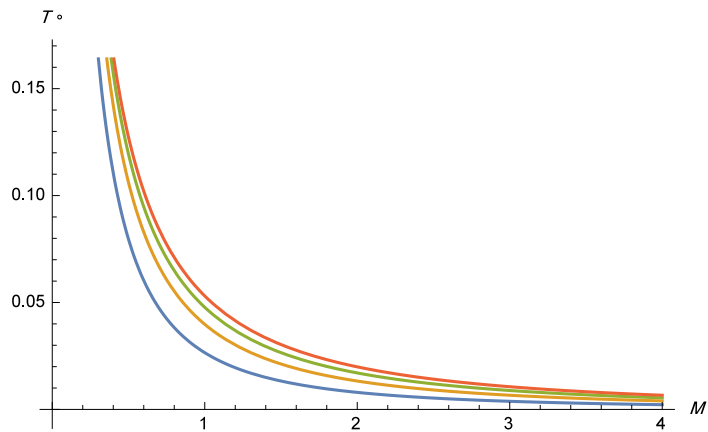}
	\caption{ Hawking temperature of different rotating K-K black holes( fixed $Q,P$ and $J$) as a function of  their mass $M$.}
	\label{T0}
\end{figure}
Using the relation $ dS = dM/T $ we can obtain the entropy:
\begin{equation}
S_0=\frac{2 \pi  \frac{p+q)}{4} \left(\frac{3(p+q)}{4\sqrt{1-j^2}}+12 \mu +\frac{pq}{\mu }\right)}{3 \hbar }
\label{szero}
\end{equation}
\begin{figure}[h]
	\centering
	\includegraphics[scale=1]{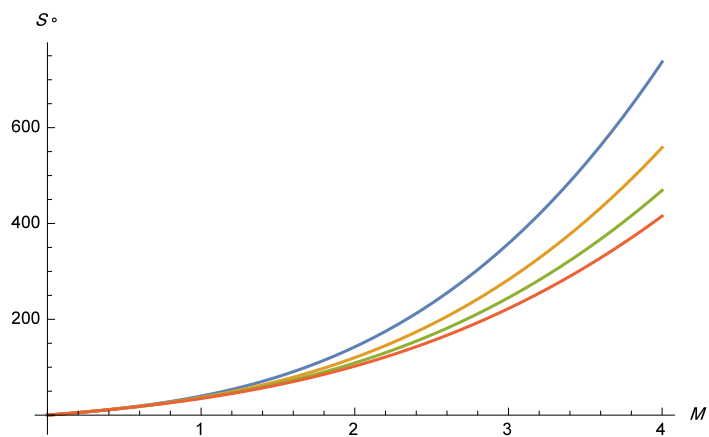}
	\caption{ The Entropy of different rotating K-K black holes( fixed $Q,P$ and $J$) as a function of  their mass $M$.}
	\label{S0}
\end{figure}
We observe ,from the figures \ref{T0} and \ref{S0}, that the K-K black holes have a very similar thermodynamic behaviour as a Kerr-Neumann black holes with an effective $ U(1)$ charge $ \mathcal Q = Q+P$.
The first law of thermodynamics is then written as \cite{larsen2000rotating}
\begin{equation}
dM= TdS - \Omega_dJ + \Phi_E dQ + \Phi_M dP
\end{equation}
In which:
\begin{align}
\Omega &= \frac{p+q}{\sqrt{pq}}\, \frac{2 \mu j}{2\mu (p+q)+(pq+4 \mu^2)\sqrt{1-j^2}}\\
\Phi_E &= \frac{\left(\frac{2 \mu }{\sqrt{1-j}}+p\right) \sqrt{\frac{p \left(q^2-\mu ^2\right)}{p+q}}}{2 \sqrt{\text{pq}} \left(\frac{2 \mu }{\sqrt{1-j^2}}+\frac{4 \mu ^2+\text{pq}}{p+q}\right)} \\
\Phi_M &= \frac{\left(\frac{2 \mu }{\sqrt{1-j}}+q\right) \sqrt{\frac{q \left(p^2-\mu ^2\right)}{p+q}}}{2 \sqrt{\text{pq}} \left(\frac{2 \mu }{\sqrt{1-j^2}}+\frac{4 \mu ^2+\text{pq}}{p+q}\right)}
\end{align}
The heat capacities can be calculated from the general relation $C_X := T(\partial S/\partial T)_X$ \cite{cai2006mass}
\begin{multline}
C_J = \frac{\pi }{2 \mu  \hbar } \Bigg(-\frac{ \mu  \hbar  \left(\sqrt{1-j^2} p+q+4\mu \right)}{\pi  \left(\sqrt{1-j^2} \mu ^2+\sqrt{1-j^2}\frac{1}{4} pq+\frac{1}{2} \mu  (p+q)\right)}\\-\left(\frac{2 \mu }{\sqrt{1-j^2}}+q\right) \sqrt{\frac{q \left(p^2-\mu ^2\right)}{p+q}}+\left(\frac{2 \mu }{\sqrt{1-j^2}}+p\right) \left(-\sqrt{\frac{p \left(q^2-\mu ^2\right)}{p+q}}\right)\Bigg).
\end{multline}
\begin{multline}
 C_Q = \frac{\pi}{2 \mu ^2 \hbar ^2}  \Bigg(-\frac{4 \mu ^2 \hbar ^2 \left(\sqrt{1-j^2} M+\mu \right)}{\pi  \left(\sqrt{1-j^2} \mu ^2+\sqrt{1-j^2} M^2+2 \mu  M\right)}- \\\frac{2 \pi ^2 \left(\sqrt{1-j^2} \mu ^2+\sqrt{1-j^2} M^2+2 \mu  M\right)^3 \left(\sqrt{1-j^2} q+2 \mu \right) \sqrt{\frac{q \left(p^2-\mu ^2\right)}{p+q}}}{\left(1-j^2\right)^{3/2} \mu  \hbar  \left(\sqrt{1-j^2} M+\mu \right)} \\- \mu  \hbar  \left(\frac{2 \mu }{\sqrt{1-j^2}}+q\right) \sqrt{\frac{q \left(p^2-\mu ^2\right)}{p+q}}\Bigg).
\end{multline}
\begin{multline}
C_P = \frac{\pi}{2 \mu ^2 \hbar ^2} \Bigg(-\frac{4 \mu ^2 \hbar ^2 \left(\sqrt{1-j^2} M+\mu \right)}{\pi  \left(\sqrt{1-j^2} \mu ^2+\sqrt{1-j^2} M^2+2 \mu  M\right)}-\\\frac{2 \pi ^2 \left(\sqrt{1-j^2} \mu ^2+\sqrt{1-j^2} M^2+2 \mu  M\right)^3 \left(\sqrt{1-j^2} q+2 \mu \right) \sqrt{\frac{q \left(p^2-\mu ^2\right)}{p+q}}}{\left(1-j^2\right)^{3/2} \mu  \hbar  \left(\sqrt{1-j^2} M+\mu \right)} \\-\mu  \hbar  \left(\frac{2 \mu }{\sqrt{1-j^2}}+p\right) \sqrt{\frac{p \left(q^2-\mu ^2\right)}{p+q}}\Bigg).
\end{multline}
Hereby, we finish the review on rotating K-K black holes. In the next section, they shall be deformed by gravity's rainbow and their thermodynamic properties will be discussed. 
\section{K-K black holes in gravity's rainbow}
\begin{figure}[h!]
	\centering
	\includegraphics[scale= 0.35]{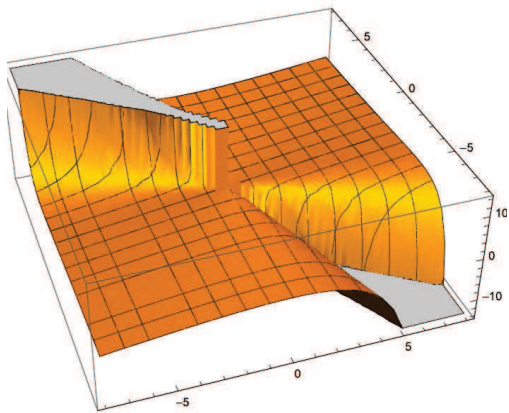}
	\caption{A plot of $G_0(T, J,,Q,P)$ of a critical rotating K-K black hole fixed $ Q,P$. Showing the critical phenomena.}
	\label{Gibbs3D}
\end{figure}

The rotating K-K black hole is deformed by the rainbow functions discussed earlier in \eqref{rainbowfunctions}, where $E$ is the energy of a ` quantum' particle near the outer horizon $ \hat r \sim r_+$. Since the K-K black hole is 4-dimensional, since the fifth dimension is compactified and  the motion on it resembles the $U(1)$ charge, the particle could - for instance- be emitted from the Hawking radiation, and this has been studied for other black holes.
\cite{ali2014black}. In order to estimate $E$, we may use the uncertainty relation for position and momentum , and write $\Delta p \geq 1/\Delta x $. Thus, we can obtain   a bound
on energy of a black hole, $E \geq 1/\Delta x $ \cite{ali2015remnant}. It should be noted that this uncertainty relation holds for the rotating K-K black hole like any other 4-D black hole, in gravity's rainbow \cite{ali2014black}. Thus we write,
\begin{equation}
E\geq 1/{\Delta x} \approx 1/{r_+}.
\end{equation} 
The general relation for temperature of a black hole in gravity rainbow was found to be \cite{ali2015remnant} :
\begin{equation}
T= T_0 \frac{g(E)}{f(E)}
\label{Trainbow}
\end{equation}
Where $f(E)$ and $ g(E)$ are the rainbow function defined in \eqref{rainbowfunctions}.Observe that these deformations depend on the radial coordinates $ \hat r$. \\
The deformation relation \eqref{Trainbow} is explained thoroughly in the following references \cite{ali2015gravitational,ali2015remnant,angheben2005hawking,ali2014black} and many others. It is natural therefore to conjecture that this deformation holds for the rotating K-K black holes, as well. 
One may define the rainbow functions $ f(E)$ and $g(E)$ in many ways, However, in this study these functions are chosen such that they are compatible with loop quantum gravity and non-commutative geometry \cite{amelino2013quantum,jacob2010modifications}.
\begin{align}
f(E) &:= 1 & g(E) &:= \sqrt{1-\eta(E/E_p)^ \nu},
\label{rainbowfunctions1}
\end{align}
 Here, $ \eta $ and $ \nu$ are free parameters.
Now, we use \eqref{Trainbow}\eqref{tzero}, and \eqref{rainbowfunctions1} to obtain the formula for the modified  temperature : 
\begin{equation}
T =  \frac{\mu  \hbar \sqrt{1-\eta(1/r_+E_p)^ \nu}  }{\pi  \sqrt{p q} \left(\frac{2 \mu }{\sqrt{1-j^2}}+\frac{4 \mu ^2+p q}{p+q}\right)}
\label{trainbow}
\end{equation}
Since the area of the 4-D black hole is spherically symmetric \cite{cai2006mass}, we have $ A= 4 \pi r^2_+$ we may re-write \eqref{Trainbow} in terms of $A$ instead of $r_+$ :
\begin{equation}
T(M) =  \frac{\mu  \hbar \sqrt{1-\eta(1/\sqrt{\frac{4 \pi}{A}}E_p)^ \nu}  }{2\pi  M \left(\frac{2 \mu }{\sqrt{1-j^2}}+\frac{4 \mu ^2+4M^2}{M}\right)}
\label{trainbow1}
\end{equation}
\begin{figure}[h]
	\centering
	\includegraphics[scale=1]{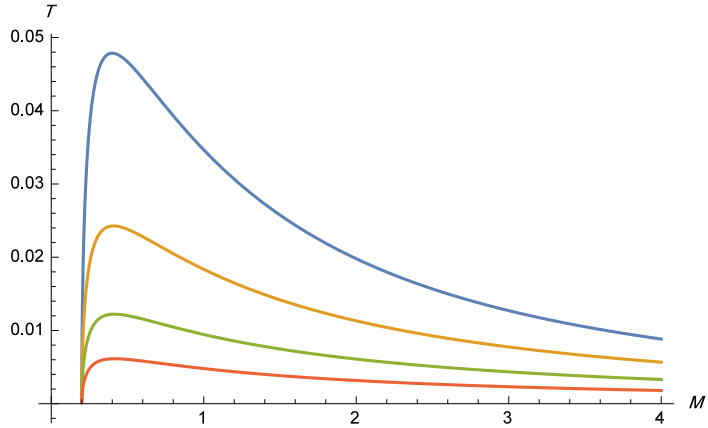}
	\caption{Deformed Hawking temperature of different rotating K-K black holes( fixed $Q,P$ and $J$) as a function of  their mass $M$. We set $ E_p=5$, $ \eta =1$ and $\nu=2$. The remnant can be observed at the same point for all black holes.}
	\label{TR}
\end{figure}
Similarly, the deformed entropy is calculated from the integral $ S= \int \frac{dM}{T}$, it is found to be given by the Hypergeometric functions $ _2F_1(a,b;c;d)$,
\begin{multline}
S(M) = \frac{2 \pi }{\mu  \hbar } \Bigg(\mu  M \Bigg(\frac{M \, _2F_1\left(\frac{1}{2},-\frac{2}{\nu};\frac{\nu-2}{\nu};\left(\frac{1}{M E_p}\right)^\nu \eta \right)}{\sqrt{1-j^2}}\\ +\mu  \, _2F_1\left(\frac{1}{2},-\frac{1}{\nu};\frac{\nu-1}{\nu};\left(\frac{1}{M E_p}\right)^\nu \eta \right)\Bigg)+\frac{1}{3} M^3 \, _2F_1\left(\frac{1}{2},-\frac{3}{\nu};\frac{\nu-3}{\nu};\left(\frac{1}{M E_p}\right)^\nu \eta \right)\Bigg)
\label{srainbow}
\end{multline}
\begin{figure}[h!]
	\centering
	\includegraphics[scale=1]{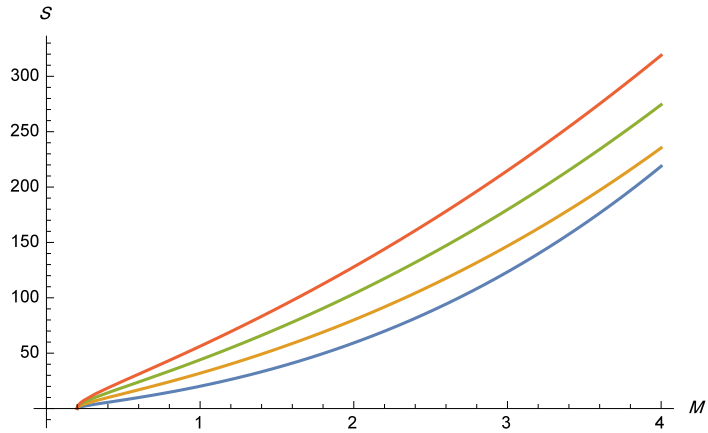}
	\caption{ The deformed entropy of different rotating K-K black holes( fixed $Q,P$ and $J$) as a function of  their mass $M$.We set $ E_p=5$, $ \eta =1$ and $\nu=2$. The remnant can be observed at the same point for all black holes.}
	\label{SR}
\end{figure}
We observe from the figures \ref{TR} and \ref{SR} the existence of a remnant, like the other studied types of deformed black holes in gravity's rainbow \cite{ali2014black}. 
The heat capacity at constant $J$ is deformed in the following way:
\begin{align}
C_J' = \frac{1}{ \sqrt{1-\eta(E/E_p)^ \nu}} C_J ,
\end{align}
same goes for other heat capacities.
It is interesting to look at the criticality of rotating K-K black holes and their rainbow deformation, this can be done by studying the Gibbs free energy of this black hole. The Gibbs free energy is generally given by:
\begin{equation}
G(M,J, Q,P) = M- TS
\end{equation}
For the ordinary rotating K-K black hole it is found to be
\begin{equation}
G_0 (M) = \frac{M^2 \left(2 \sqrt{1-j^2} M+3 \mu \right)}{3 \left(\sqrt{1-j^2} \mu ^2+\sqrt{1-j^2} M^2+2 \mu  M\right)}.
\label{Gibbs}
\end{equation}
We can Plot \eqref{Gibbs}, keeping $,Q,P$ fixed and vary $M$ and $J$ to obtain the plot \ref{Gibbs3D} , that shows a critical phenomena for the rotating K-K black holes.
\begin{figure}[h!]
	\centering
	\includegraphics[scale= 0.4]{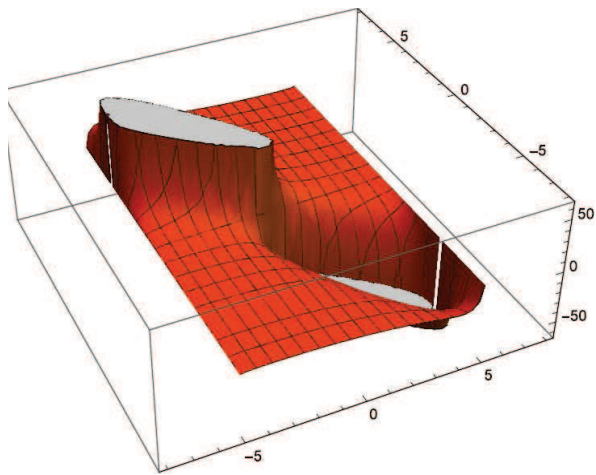}
	\caption{A plot of the deformed Gibbs free energy $G(T, J,,Q,P)$ of a deformed rotating K-K black hole fixed $ Q,P$. Showing the  same critical phenomena, as the ordinary K-K black hole. We have set $ \eta =1,  E_p=5$ and $ \nu=2$ .}
	\label{GibbsR3D}
\end{figure}
The deformed Gibbs free energy is calculated from  \eqref{SR} and \eqref{TR},
\begin{multline}
G = M-\frac{\sqrt{1-\eta  \left(\frac{M}{E_p}\right)^{\nu }}}{\frac{2 \mu  M}{\sqrt{1-j^2}}+\mu ^2+M^2} \Bigg(\frac{\mu  M^2 \, _2F_1\left(\frac{1}{2},\frac{2}{\nu };\frac{\nu +2}{\nu };\left(\frac{M}{E_p}\right)^{\nu } \eta \right)}{\sqrt{1-j^2}} \\ +\frac{1}{3} M^3 \, _2F_1\left(\frac{1}{2},\frac{3}{\nu };\frac{\nu +3}{\nu };\left(\frac{M}{E_p}\right)^{\nu } \eta \right)+\mu ^2 M \, _2F_1\left(\frac{1}{2},\frac{1}{\nu };1+\frac{1}{\nu };\left(\frac{M}{E_p}\right)^{\nu } \eta \right)\Bigg)
\end{multline}
Both ordinary and deformed rotating K-K black holes show critical behaviour as the study of Gibbs free energy, if $ G>0$ the black hole is said to be ` critical' and when $ G<0$ it is said that the black hole is uncritical.  
\section{Conclusion}
In this paper, the geometry of 5-D rotating Kaluza Klein black holes with electric and magnetic charges was deformed by the rainbow functions $F,G$ motivated by loop quantum gravity and non-commutative geometry. Resulting a deformation on the thermodynamics of the 4D rotating K-K black hole. The  deformed temperature and entropy indicate the existence of a remnant after the decay of the black hole to a ` Plankckian' scale.  This is independent fro the compactification, or the K-K reduction of the 5-D geometry.  Moreover, the critical behaviour of this black hole was studied via calculating its Gibbs free energy, the ordinary and the deformed black holes appear to show the same critical behaviour.
\section*{Acknowledgements}
{ \fontfamily{times}\selectfont
	\noindent 
	Warm regards to Dr Mir Faizal for his generous help improving this work. \\ This research project was supported by a grant from the " Research Center of the Female Scientiffic and Medical Colleges ", Deanship of Scientiffic Research, King Saud University. 
\bibliography{ref}
\bibliographystyle{plain}
\end{document}